\documentclass[
 aps, pra,
 amsmath,amssymb,
 11pt,
 final,
tightenlines,
 twoside,
 twocolumn,
 nofloats,
nofootinbib,
 superscriptaddress,
showkeys,
showkeywords,
 ]
{revtex4-2}

\usepackage[T1]{fontenc}
\usepackage[utf8x]{inputenc}
\usepackage[english]{babel}
\usepackage{graphicx}
\usepackage{dcolumn}
\usepackage{bm}
\usepackage{subfigure}
\usepackage{url}

\input{maik.rty}

\setcitestyle{authoryear,round}
\def\squareforqed{\hbox{\rlap{$\sqcap$}$\sqcup$}}

\def\sq{\ifmmode\squareforqed\else{\unskip\nobreak\hfil
\penalty50\hskip1em\null\nobreak\hfil\squareforqed
\parfillskip=0pt\finalhyphendemerits=0\endgraf}\fi}

\def\utw{\smash{\rlap{\lower5pt\hbox{$\sim$}}}}

\def\udtw{\smash{\rlap{\lower6pt\hbox{$\approx$}}}}

\def\diameter{{\ifmmode\mathchoice
{\ooalign{\hfil\hbox{$\displaystyle/$}\hfil\crcr
{\hbox{$\displaystyle\mathchar"20D$}}}}
{\ooalign{\hfil\hbox{$\textstyle/$}\hfil\crcr
{\hbox{$\textstyle\mathchar"20D$}}}}
{\ooalign{\hfil\hbox{$\scriptstyle/$}\hfil\crcr
{\hbox{$\scriptstyle\mathchar"20D$}}}}
{\ooalign{\hfil\hbox{$\scriptscriptstyle/$}\hfil\crcr
{\hbox{$\scriptscriptstyle\mathchar"20D$}}}}
\else{\ooalign{\hfil/\hfil\crcr\mathhexbox20D}}%
\fi}}

\newcommand{\obj}{V379~Vir}

\begin{document}

\selectlanguage{english}



\title{Phase-Resolved Spectroscopy of the Polar V379 Vir with a Brown Dwarf}

\author{\firstname{M.~V.}~\surname{Suslikov}}
 \email{mvsuslikov@outlook.com}
 \affiliation{Kazan Federal University, Kazan, 420008 Russia}
 \affiliation{Special Astrophysical Observatory,  Russian Academy of Sciences, Nizhnii Arkhyz, 369167 Russia}

\author{\firstname{A.~I.}~\surname{Kolbin}}
 \affiliation{Special Astrophysical Observatory,  Russian Academy of Sciences, Nizhnii Arkhyz, 369167 Russia}

 \author{\firstname{N.~V.}~\surname{Borisov}}
 \affiliation{Special Astrophysical Observatory,  Russian Academy of Sciences, Nizhnii Arkhyz, 369167 Russia}

\begin{abstract}
The polar {\obj} is a well-known magnetic cataclysmic variable with a brown dwarf donor. Despite numerous studies of this system across various spectral ranges, a detailed investigation of the orbital variability of its optical spectra has not been carried out. In this work, we present an analysis of spectroscopic observations of {\obj} obtained with the 6-m telescope of the Special Astrophysical Observatory of the Russian Academy of Sciences. The orbital variability of the H$\alpha$ emission indicates that the line is most likely formed in the accretion stream near the Lagrangian point L$_1$, rather than on the donor's surface as previously assumed. The analysis of the rotational variability of the Zeeman splitting of hydrogen lines reveals a complex magnetic field topology of the white dwarf, which differs from a simple dipole configuration.
\end{abstract}

\maketitle

\section{INTRODUCTION}
\label{Introduction}
\noindent

Polars (or AM~Her-type stars) are a subclass of cataclysmic variables --- semi-detached binaries consisting of an accreting white dwarf and a low-mass donor star \citep{Warner1995, Hellier2001}. In these systems, the white dwarf possesses a strong magnetic field ($B \sim 10-100$~MG), which prevents the formation of an accretion disk by channeling the accretion flow along magnetic field lines toward one or both magnetic poles \citep{Cropper1990}. When the infalling gas impacts the surface of the white dwarf, accretion spots are formed. These spots are sources of X-ray emission and polarized cyclotron radiation observed in the optical and infrared ranges. Due to the strong magnetic field, polars are typically synchronous systems (except for a small group of so-called asynchronous polars) in which the white dwarf's spin period is locked to the orbital period.

The polar {\obj} (also known as SDSS J121209.31+013627.7) has been the subject of numerous studies. The unique nature of {\obj} was first noted by \cite{Schmidt2005}, who showed that this system is a short-period ($P_\mathrm{orb} = 88.4$~min) cataclysmic variable with a magnetic white dwarf ($B \approx 7$~MG) and an L-type brown dwarf companion. The source belongs to a relatively rare group of evolved cataclysmic variables known as period bouncers \citep{Knigge2011}. \citet{Schmidt2005} reported a very low accretion rate in {\obj} and suggested that the system might be detached, with accretion proceeding through the stellar wind from the secondary. Near-infrared observations confirmed the presence of a cool donor with an effective temperature of $T_{\mathrm{eff}} \le 1700$~K and revealed variable cyclotron emission \citep{Debes2006, Farihi2008}. An analysis of XMM-Newton X-ray observations showed spin-modulated variability caused by the presence of an accretion spot on the white dwarf's surface \citep{Stelzer2017}. Based on spectral energy distribution (SED) modeling, \citet{Suslikov2025} estimated the white dwarf's mass as $M_1 = 0.61 \pm 0.05~M_{\odot}$, its effective temperature as $T_{\mathrm{eff}} = 10930 \pm 350$~K, and derived an accretion rate of $\dot{M} \approx 3 \times 10^{-13}~M_{\odot}/\mathrm{yr}$. It was suggested that {\obj} is a semi-detached binary system experiencing a prolonged ($\approx 20$ year) low state. In this paper, we present the results of a phase-resolved spectroscopic study of {\obj}. Our aim is to investigate the orbital variability of the H$\alpha$ emission line, which carries information about the structure of the accretion flow, and to examine the rotational modulation of the white dwarf's magnetic field.

\section{OBSERVATIONS AND DATA REDUCTION}
\label{Observations and data reduction}
\subsection{Spectroscopy}
\label{Spectroscopy}

Spectroscopic observations of {\obj} were obtained with the 6-m BTA telescope at the Special Astrophysical Observatory of the Russian Academy of Sciences (SAO RAS). The telescope was equipped with the SCORPIO-1 focal reducer\footnote{A description of the SCORPIO-1 instrument can be found at https://www.sao.ru/hq/lsfvo/devices/scorpio /scorpio.html.}, operated in the long-slit spectroscopy mode \citep{Afanasiev2005}. The observations were conducted in two sets: during the night of 7/8 May 2007 and the night of 25/26 April 2022. In the first run, the VPHG1200G grism (1200 lines/mm) and a slit width of $1''$ were used, providing spectra in the 4000--5700~\AA~range with a spectral resolution of $\Delta \lambda \approx 5$~\AA. During the second run, the VPHG550G grism (550 lines/mm) and a slit width of $1.2''$ were employed, yielding spectra covering 4000--7300~\AA~with a resolution of $\Delta \lambda \approx 12$~\AA. In total, 15 and 22 spectra of {\obj} were obtained during the first and second observing sets, respectively, covering the full orbital period of the system. Each spectrum was acquired with an exposure time of 300~s. The log of the observations is presented in Table \ref{tab:log_phot}.

The spectroscopic data were processed following standard long-slit reduction techniques implemented in the IRAF software package\footnote{The IRAF software package is available at https://iraf-community.github.io.}. The raw images were bias-subtracted and cleaned of cosmic rays using the L.A.Cosmic algorithm \citep{Dokkum2001}. Flat-fielding was applied using exposures of a continuum lamp. The correction for geometric distortions and the wavelength calibration were carried out using frames of a He-Ne-Ar arc lamp. The spectra were optimally extracted \citep{Horne1986} with sky background subtraction. Spectrophotometric calibration was performed using observations of the standard stars Hz44 (for the 2007 data) and AGK+81$^{\circ}$266 (for the 2022 data). For each spectrum, we computed the barycentric Julian date at mid-exposure and the barycentric velocity correction. The orbital phases corresponding to the observation times were calculated using the orbital period $P_\mathrm{orb}$ from \cite{Suslikov2025} and the reference epoch $BJD(\varphi=0) = 2453798.5956(5)$, derived under the assumption that the H$\alpha$ emission originates from the surface of the donor star (see Section \ref{Doppler tomography}).

\begin{table*}
    \caption{Observation log for {\obj}. The table lists the telescopes and instruments used, the observing nights, the total observing time, the number of acquired spectra ($N$), the covered spectral range, and the exposure time ($\Delta t_{\mathrm{exp}}$).}
    \label{tab:log_phot}
    \vspace*{1mm}
    \begin{center}
        \begin{tabular}{lccccc}
            \hline
            Telescope/      & Date              & Duration & $N$ & Spectral range & $\Delta t_{\mathrm{exp}}$ \\
            Instrument      & (LT)              & (min)    &     &                & (s)                       \\ \hline
            BTA/SCORPIO-1   & 07/08 May 2007    & 86       & 15  & 3900--5700~\AA & 300                       \\
            BTA/SCORPIO-1   & 25/26 Apr. 2022   & 118      & 22  & 4000--7200~\AA & 300                       \\
            RTT-150/TFOSC   & 07/08 May 2022    & 182      & 72  & $B$            & 120                       \\
            \hline
        \end{tabular}
    \end{center}
    \vspace*{-5mm}
\end{table*}

\subsection{Photometry}
\label{Photometry}

Photometric observations of {\obj} were performed on the night of 7/8 April 2022 using the 1.5-m RTT-150 telescope at the T\"UB\.ITAK National Observatory (Turkey) with the TFOSC instrument in imaging mode. A total of 72 direct images were obtained in the Johnson B band, each with an exposure time of 120 s. The photometric data were reduced using the astropy\footnote{The astropy package is available at https://www.astropy.org.} and photutils\footnote{The photutils package is available at https://photutils.readthedocs.io.} Python libraries. The preprocessing included subtraction of a median bias frame, flat-field correction, and removal of cosmic-ray hits. PSF photometry was applied to maximize the signal-to-noise ratio.

\section{ANALYSIS OF THE H$\alpha$ EMISSION}
\label{H_alpha analysis}
\subsection{Radial velocities}
\label{Radial velocities}

\begin{figure*}[]
    \centering 
    \includegraphics[width=\textwidth]{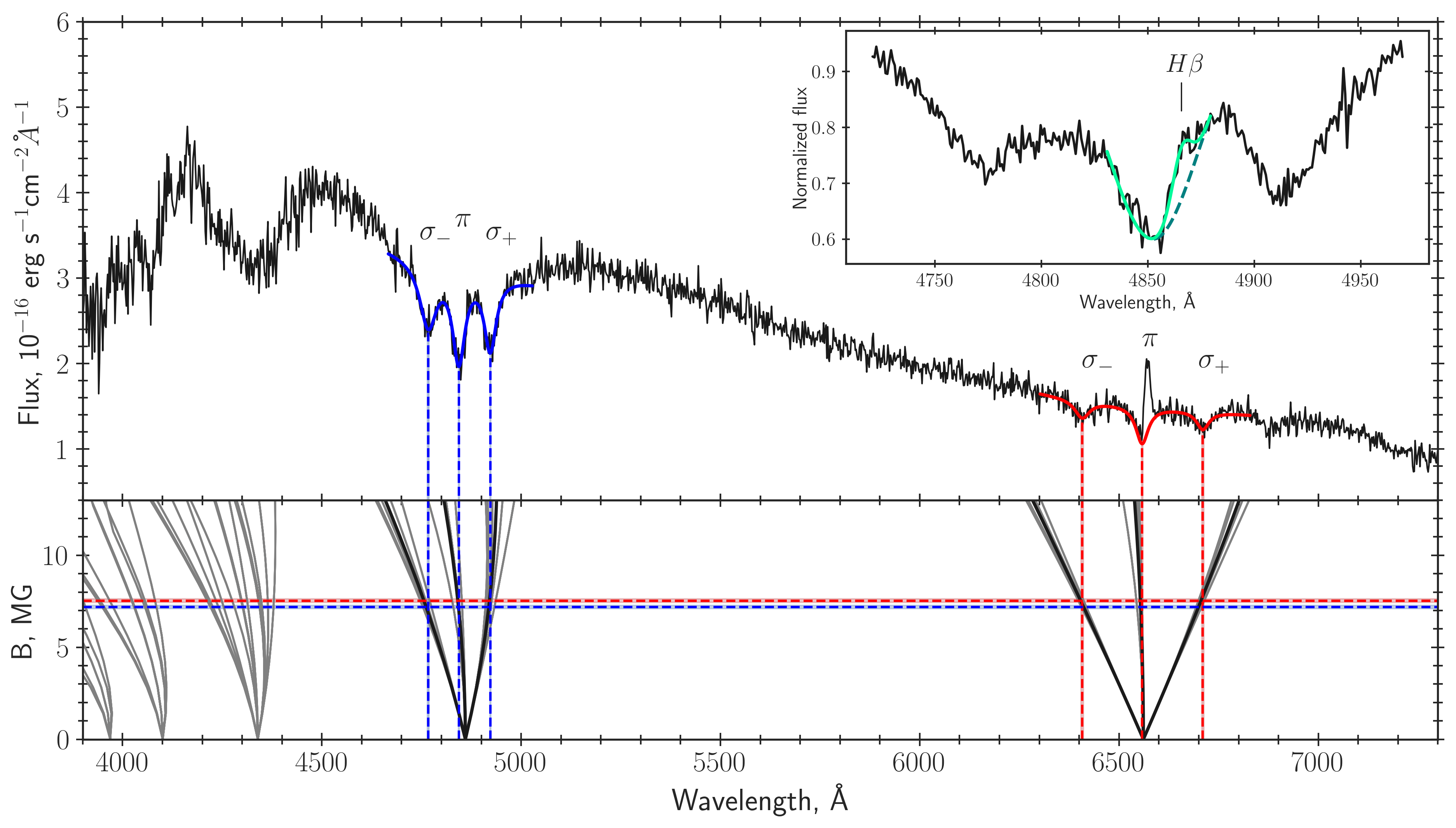} 
    \caption{Example spectrum of {\obj} (top panel) and a diagram of the Zeeman splitting of the hydrogen lines (bottom panel). The Zeeman triplets of H$\alpha$ and H$\beta$ are modeled using a set of Lorentzian profiles. Vertical lines indicate the positions of the triplet components. The inset in the top panel shows the normalized averaged spectrum in the region of the H$\beta$ line, revealing weak emission.}
    \label{fig:spec_example}
\end{figure*}

\begin{figure*}[]
    \centering 
    \vspace*{3mm}
    \includegraphics[width=\textwidth]{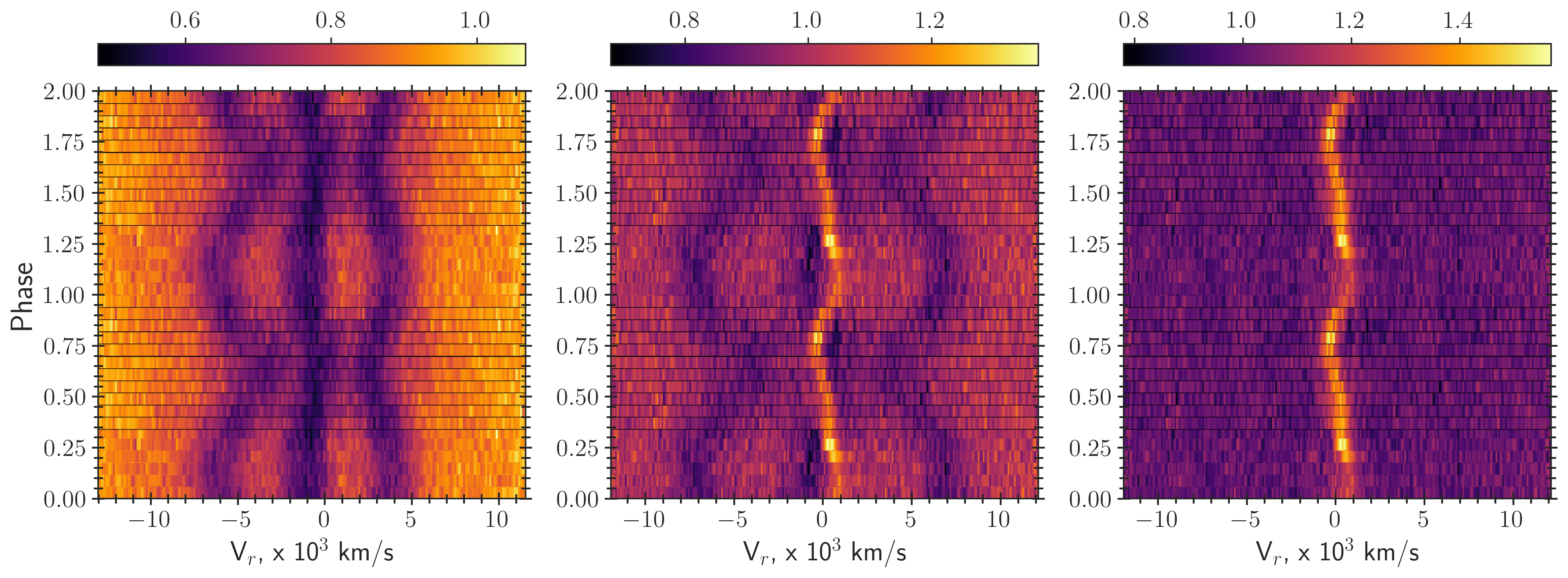} 
    \vspace*{-7mm}
    \caption{Dynamic spectra in the regions of the H$\beta$ (left panel) and H$\alpha$ lines before (central panel) and after (right panel) subtraction of the absorption background.}
    \label{fig:dynamic_specs}
\end{figure*}

The spectral series obtained with the BTA telescope exhibit a blue continuum slope, the presence of absorption components of the Zeeman-split H$\alpha$, H$\beta$, H$\gamma$, and H$\delta$ lines, as well as weak H$\alpha$ emission (see Fig.~\ref{fig:spec_example}). The phase-averaged 2007 spectrum also reveals weak H$\beta$ emission superimposed on the Zeeman components. Overall, the spectral features of {\obj} are consistent with those expected for a weakly accreting magnetic white dwarf, as previously noted by \cite{Schmidt2005}. The Zeeman components shift periodically due to changes in the orientation of the magnetic field relative to the observer. The position of the weak H$\alpha$ emission line varies over the orbital period. Dynamic spectra in the regions of the H$\beta$ and H$\alpha$ lines, reflecting the periodic variations in the line components, are shown in Fig.~\ref{fig:dynamic_specs}.

To analyse the H$\alpha$ emission, it is necessary to separate it from the absorption $\pi$ and $\sigma^{\pm}$ components, which shift over the orbital period. For this purpose, the local continuum around the H$\alpha$ line was first subtracted using a low-order polynomial. The absorption components were then modeled as a sum of three Gaussians $A_i \exp[-(\lambda-\lambda^0_i)^2/2 \sigma^2_i]$, where the index $i$ corresponds to the splitting component ($i=\pi, \sigma^-$, $\sigma^+$), $A$ is the Gaussian amplitude, $\sigma$ is the standard deviation, and $\lambda^0$ is the central wavelength of the Zeeman component. Spectral regions containing the H$\alpha$ emission were excluded from the fit. The Gaussian parameters (amplitude, central wavelength, and width) were assumed to vary with orbital phase according to a harmonic law $f_j(\varphi) = a_{i,j} + b_{i,j} \sin[2\pi (\varphi-\varphi_{i,j}^0)]$, where $i = \pi$, $\sigma^-$, $\sigma^+$, and the index $j$ corresponds to the Gaussian parameter ($j = A$, $\lambda^0$, $\sigma$). The parameters of the moving Gaussians were determined via a least-squares fitting. The $\chi^2$ minimization was performed using the Nelder--Mead method. The fit converged to the dynamic spectrum of the absorption background with a reduced chi-square of $\chi^2_{\nu} = 1.33$. Dynamic spectra of the H$\alpha$ region before and after subtraction of the absorption background are shown in Fig.~\ref{fig:dynamic_specs}.

The radial velocities of the H$\alpha$ emission are poorly fitted by a single sinusoid, indicating the presence of either multiple emission sources or a single extended emitting region. It is known that the emission lines in polars have a multi-component structure \citep{Schwope1997} and, at moderate spectral resolution, can often be satisfactorily described by two components (see, e.g., \citealt{Liu2023, Kolbin2023}). To investigate the behavior of the H$\alpha$ emission in {\obj}, we modeled the line using two Gaussian profiles. The widths of the profiles were kept fixed, while their radial velocities were allowed to vary sinusoidally over the orbital period. The Gaussian amplitudes were optimized individually for each spectrum. The line component parameters (Gaussian amplitudes and widths, radial velocity semi-amplitudes, zero phases, and $\gamma$-velocities) were determined via a least-squares minimization. The dynamic spectrum is best reproduced by a combination of bright (b) and faint (w) sources, with radial velocity semi-amplitudes of $K_b = 320 \pm 13$~km/s and $K_w = 989 \pm 27$~km/s, profile widths of $\sigma_b = 391 \pm 8$~km/s and $\sigma_w = 309 \pm 14$~km/s, and $\gamma$-velocities of $\gamma_b = 120 \pm 10$~km/s and $\gamma_w = -29 \pm 18$~km/s\footnote{It should be noted that radial velocities may be uncertain by several tens of km/s due to uneven illumination of the SCORPIO-1 slit.}, respectively. The bright component is delayed in orbital phase relative to the faint one by $\Delta \varphi = 0.16 \pm 0.01$. The quality of the fit is characterized by a reduced chi-square of $\chi^2_{\nu} = 1.25$. For comparison, a single Gaussian model yields a noticeably poorer fit with $\chi^2_{\nu} = 1.63$. \cite{Schmidt2005} proposed that most of the H$\alpha$ emission originates from the donor star's surface, consistent with the semi-amplitude measured here for the bright component ($K = 320 \pm 20$ km/s). The $\gamma$-velocity of the faint source is lower by $\sim 150$ km/s compared to the bright source, suggesting that the accretion-flow material may be lifted above the orbital plane by the white dwarf's magnetic field.

\subsection{Doppler tomography}
\label{Doppler tomography}

The Doppler tomography technique allows the reconstruction of the distribution of emission line sources in velocity space from phase-resolved spectroscopy. This method typically uses a two-dimensional velocity space defined by basis vectors in the orbital plane. Each point in this space can be described by pair of polar coordinates: the velocity magnitude $\upsilon$ relative to the system's center of mass (defined up to a factor of $\sin i$, where $i$ is the orbital inclination), and the angle $\vartheta$ specifying the direction of motion of the emitting particle (usually measured from the line connecting the centers of mass of the stellar components). Doppler tomograms can be represented in two ways: the standard projection, in which the velocity magnitude $\upsilon$ increases outward from the center, and the inside-out projection, where $\upsilon$ decreases from the center toward the outer regions. The inside-out projection is particularly useful for analyzing high-velocity structures that would otherwise be spread over a large area in the standard projection. For a detailed description of the Doppler tomography method, we refer to \cite{Marsh2016, Kotze2015, Kotze2016}.

\begin{figure}[h!]
    \begin{minipage}[t]{\columnwidth}
    \vspace*{3mm}
    \includegraphics[width=0.95\linewidth]{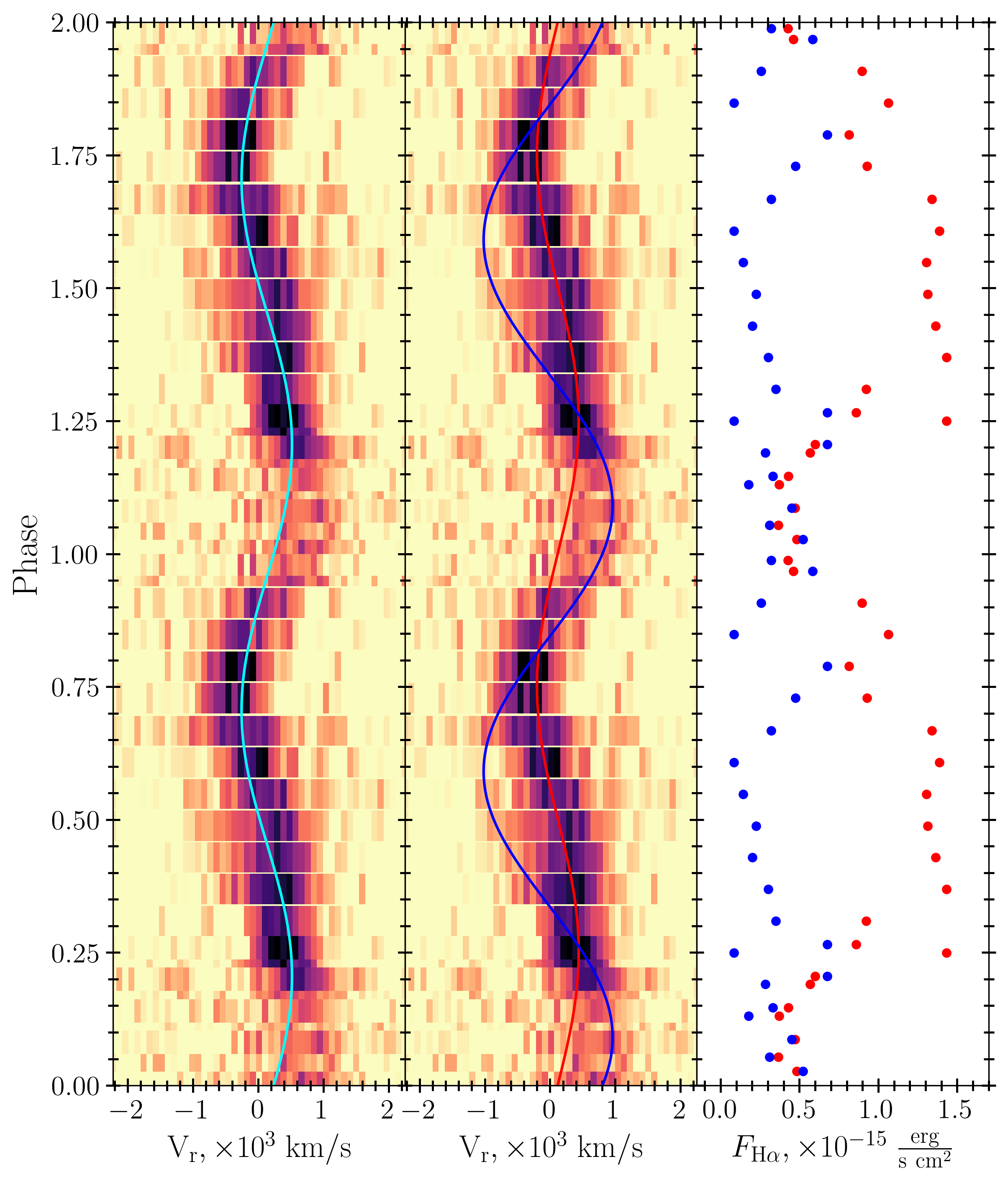}
    \vspace*{-2mm}
    \caption{Left panel: Dynamic spectrum of the H$\alpha$ emission after absorption background subtraction, overlaid with the radial velocity curve of a single source. Middle panel: The same dynamic spectrum, showing the radial velocity curves of the bright slow (red sinusoid) and faint fast (blue sinusoid) components. Right panel: Orbital modulation of the integrated H$\alpha$ flux for the bright (red points) and faint (blue points) sources.}
    \label{fig:dynspec}
    \end{minipage}
\end{figure}

\begin{figure*}%
    \centering
    \subfigure[]{\includegraphics[width=0.35\textwidth]{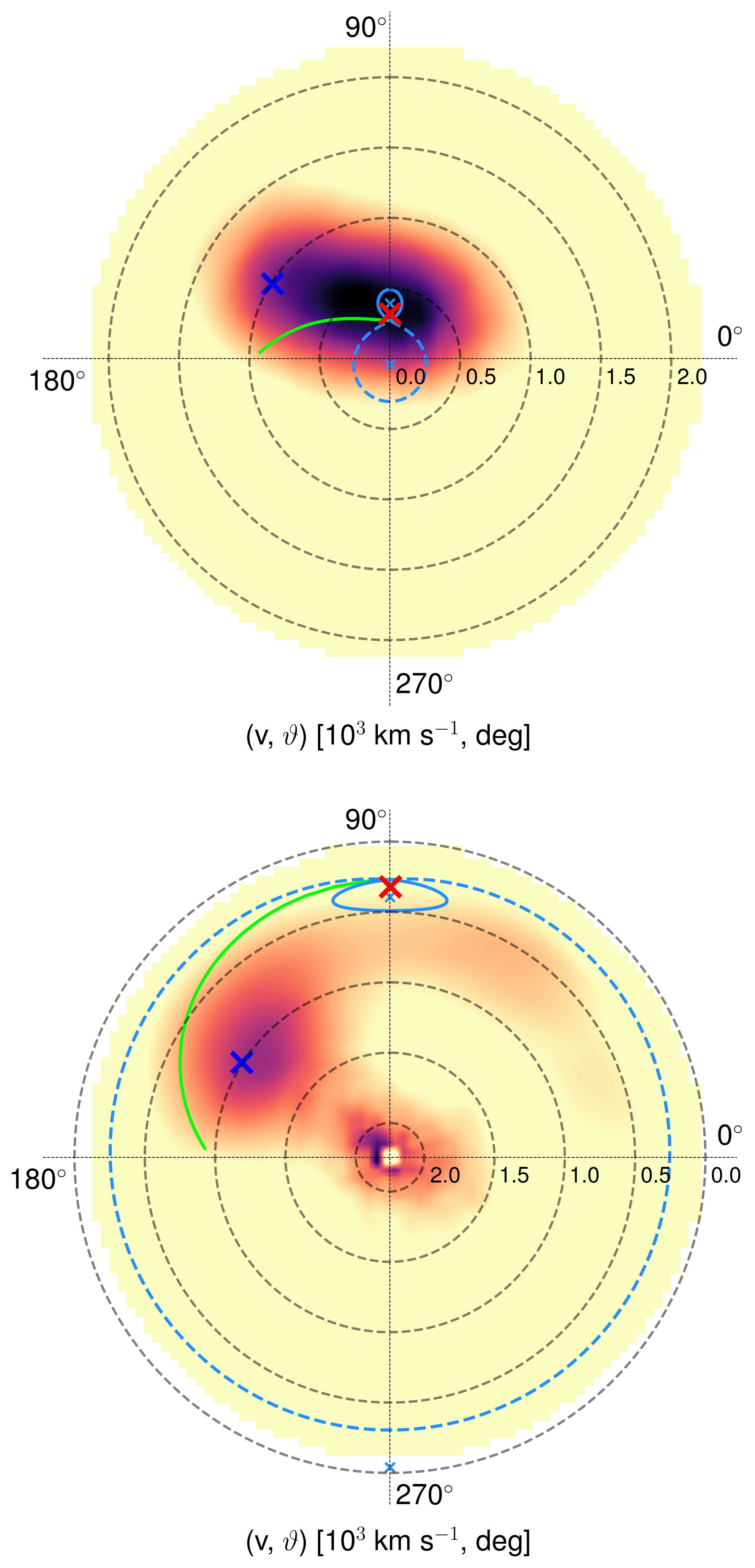} }
    \hspace{3cm} 
    \subfigure[]{\includegraphics[width=0.35\textwidth]{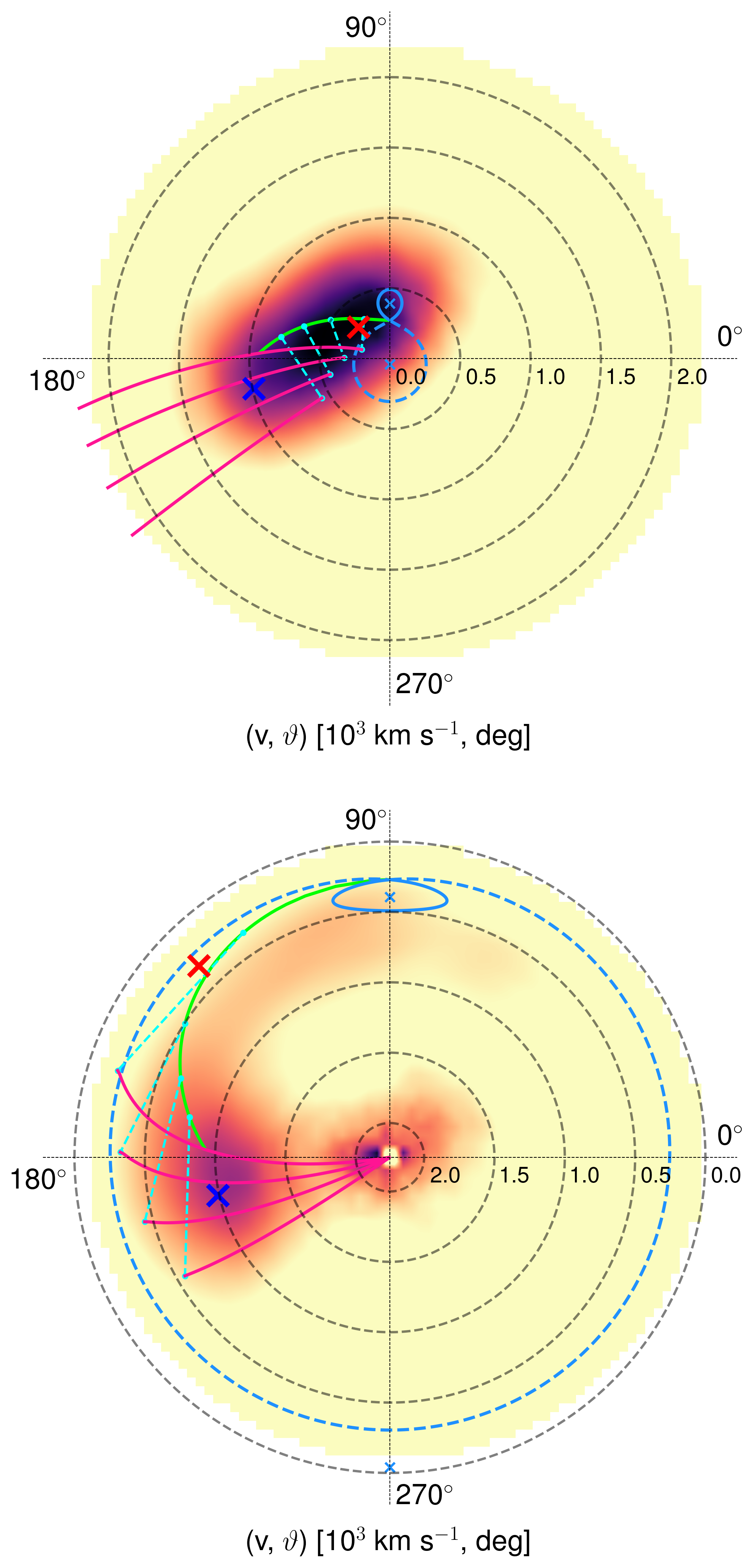} }
    \caption{Doppler tomograms of {\obj} in the H$\alpha$ emission line, reconstructed under the assumption that the bright source forms on the irradiated surface of the donor (a), and the same tomograms rotated by an angle of $\Delta \vartheta \approx 45^\circ$ (b). The emission centroids of the bright (red cross) and faint (blue cross) components are marked. The ballistic trajectory of the accretion flow is shown by the green line, and the particle velocities along the magnetic trajectory are indicated by the red lines.}
    \label{fig:doptomog}%
\end{figure*}

To reconstruct the Doppler maps of {\obj} in the H$\alpha$ line, we employed the code of \cite{Kotze2016}, which implements the maximum-entropy method. Since the H$\alpha$ emission shows variable flux (see Fig.~\ref{fig:dynspec}), caused by eclipses or the high optical depth of the emitting region, we used the flux-modulated tomographic approach, which assumes a sinusoidal variation of the source intensity over the orbital period \citep{Steeghs2003}. The orbital phases were calculated under the assumption that the bright component of the line originates on the irradiated face of the donor star \citep{Schmidt2005}. Within this framework, phase $\varphi = 0$ corresponds to the donor's inferior conjunction, where the velocity of the bright component equals $\gamma_b$. The resulting Doppler tomograms in both the standard and inside-out projections are shown in Fig.~\ref{fig:doptomog}a. The locations of the bright and faint sources, previously identified through decomposition of the dynamic spectrum using moving Gaussians, are marked on the maps. The bright component lies along the $\vartheta = 90^{\circ}$ axis, meaning its motion is aligned with that of the donor. This follows directly from our assumption that the source is located on the donor's surface. A system model based on the binary parameters derived by \cite{Suslikov2025} is overlaid on the Doppler maps. The orbital inclination of $i \approx 60^{\circ}$ was adopted to match the region of maximum H$\alpha$ intensity with the irradiated surface of the donor. The faint source exhibits higher velocities than particles following the ballistic trajectory, suggesting that it is most likely produced in the stream flowing along the magnetic field lines. However, if this is indeed the case, the location of the magnetically confined part of the stream is atypical for polars. It usually lies in the third quadrant of the tomogram or close to it (see, e.g., \citealt{Schwope1997, Kolbin2022}). One may assume that the bright component does not originate on the donor's surface but instead forms in the accretion flow near the Lagrangian point L$_1$. In this scenario, the tomogram can be rotated by an angle of $\Delta \vartheta \approx 45^\circ$ so that the bright source aligns with the ballistic trajectory (see Fig.~\ref{fig:doptomog}b). With this orientation, the position of the faint component becomes consistent with gas motion along magnetic field lines. As an illustration, we have overplotted particle trajectories following the magnetic lines on the rotated tomograms, calculated for a magnetic dipole inclination of $\beta = 25^{\circ}$ and a magnetic pole longitude of $\psi = 45^{\circ}$.

\subsection{The nature of the H$\alpha$ emission}
\label{H_alpha nature}

Two hypotheses for the formation of H$\alpha$ emission in {\obj} have been previously discussed: (i) irradiation of the donor's surface by X-rays from the accretion spot \citep{Schmidt2005}, and (ii) chromospheric activity of the donor \citep{Howell2006}, concentrated near the Lagrangian point L$_1$. Evidence that the H$\alpha$ emission is not produced by irradiation can be inferred from the fact that the flux in an H$\alpha$ line generated by reprocessing cannot exceed the X-ray flux incident on the donor's surface. This leads to the inequality $F_{\mathrm{H}\alpha}/F_X \lesssim \Omega / 4 \pi$, where $F_{\mathrm{H}\alpha} \approx 1.4\times10^{-15}$~erg~cm$^{-2}$~s$^{-1}$ is the flux in the bright component of the H$\alpha$ emission, $F_X \approx 1.5 \times 10^{-13}$~erg~cm$^{-2}$~s$^{-1}$ is the X-ray flux \citep{Stelzer2017}, and $\Omega$ is the solid angle subtended by the donor as seen from the white dwarf. Assuming that the brown dwarf fills its Roche lobe, the solid angle can be expressed as $\Omega = \pi (R_L/A)^2$, where $R_L$ is the effective radius of the donor's Roche lobe, and $A$ is the binary separation. The ratio $R_L/A$ can be expressed in terms of the binary mass ratio using the relation from \cite{Eggleton1983}. Thus, adopting a white dwarf mass of $M_1 = 0.61~M_{\odot}$ \citep{Suslikov2025}, it follows that under the irradiation scenario, the observed flux ratio is only possible if $M_2 > 0.08~M_{\odot}$. This constraint is inconsistent with the substellar nature of the donor in {\obj}, implying that the H$\alpha$ emission must have a different origin.

An estimate of the flux ratio in the Balmer lines can provide indirect insight into the emission formation mechanism. The phase-averaged integrated fluxes in the H$\alpha$ and H$\beta$ lines are $\langle F_{\mathrm{H}\alpha}\rangle = 1.27 \times 10^{-15}$~erg~cm$^{-2}$~s$^{-1}$ and $\langle F_{\mathrm{H}\beta}\rangle = 4.56 \times 10^{-16}$~erg~cm$^{-2}$~s$^{-1}$, respectively. This yields a Balmer decrement of $\langle F_{\mathrm{H}\alpha}\rangle / \langle F_{\mathrm{H}\beta}\rangle \approx 2.8$, which is typical for a fluorescence mechanism in an optically thin gas. However, we cannot definitively determine the region of emission formation based solely on the Balmer decrement due to uncertainties in the measurement of the H$\beta$ flux.

Doppler tomograms of {\obj}, examined in the previous subsection, also raise doubts about the possibility that the H$\alpha$ emission originates on the donor's surface. The primary emission source exhibits an extended structure, implying that at least part of the H$\alpha$ emission must be produced within the accretion stream. If one assumes that the bright H$\alpha$ component forms on the donor, the resulting Doppler tomograms become difficult to reconcile with the "standard" accretion model for a polar, which includes both ballistic and magnetic trajectories. In contrast, a configuration typical of polars can be reproduced by a slight rotation of the tomogram, placing the bright source within the accretion flow. The observed variations in the H$\alpha$ flux can then be interpreted as eclipses of different parts of the accretion stream by the donor.

Similar effects have been reported for the polar BM~CrB in its low state \citep{Kolbin2023}. In that system, eclipses of the emission source were observed which, at first glance, resemble the disappearance of an irradiated region behind the donor's limb. However, such an interpretation cannot reproduce either the observed eclipse width or the radial velocity curve of the emission features, indicating that the accretion stream near the donor is the most likely source of the emission. Another example is the eclipsing system 1RXS J184542.4+483134 \citep{Kochkina2023}. Its emission line profiles show a narrow component whose behavior is reminiscent of an irradiation region. Nevertheless, Doppler tomography demonstrates that this emission component forms at a considerable distance from the donor.

\section{MAGNETIC FIELD OF THE WHITE DWARF}
\label{Magnetic field of WD}
\subsection{Magnetic curve}
\label{Magnetic curve}

As noted in the previous section, the spectra of {\obj} contain absorption components of the Zeeman-split H$\alpha$ and H$\beta$ lines, whose wavelengths vary with the orbital period and, consequently, with the white dwarf's spin period. To determine the positions of the $\sigma^{\pm}$ and $\pi$ components, their spectral region was modeled as the sum of three Lorentzian profiles, $f(\lambda) = A \gamma^2 / \left[\gamma^2 + (\lambda - \lambda_0)^2\right]$, where $\lambda_0$ is the central wavelength of the splitting component, and $A$ and $\gamma$ are free parameters. The H$\alpha$ emission was removed from the spectra prior to fitting. Uncertainties in the component central wavelengths were estimated using a Monte Carlo method. To measure the magnetic field of the white dwarf, we calculated the energies of allowed transitions for a hydrogen atom in a strong magnetic field. The energy spectrum was computed using the code described by \cite{Schimeczek2014}. The wavelengths of the transitions to the second energy level corresponding to the H$\alpha$ and H$\beta$ lines for magnetic field in the range $B = 0-13$~MG are shown in the bottom panel of Fig.~\ref{fig:spec_example}.

\begin{figure}[h!]
    \centering
    \begin{minipage}[t]{\columnwidth}
    \centering
    \includegraphics[width=\linewidth]{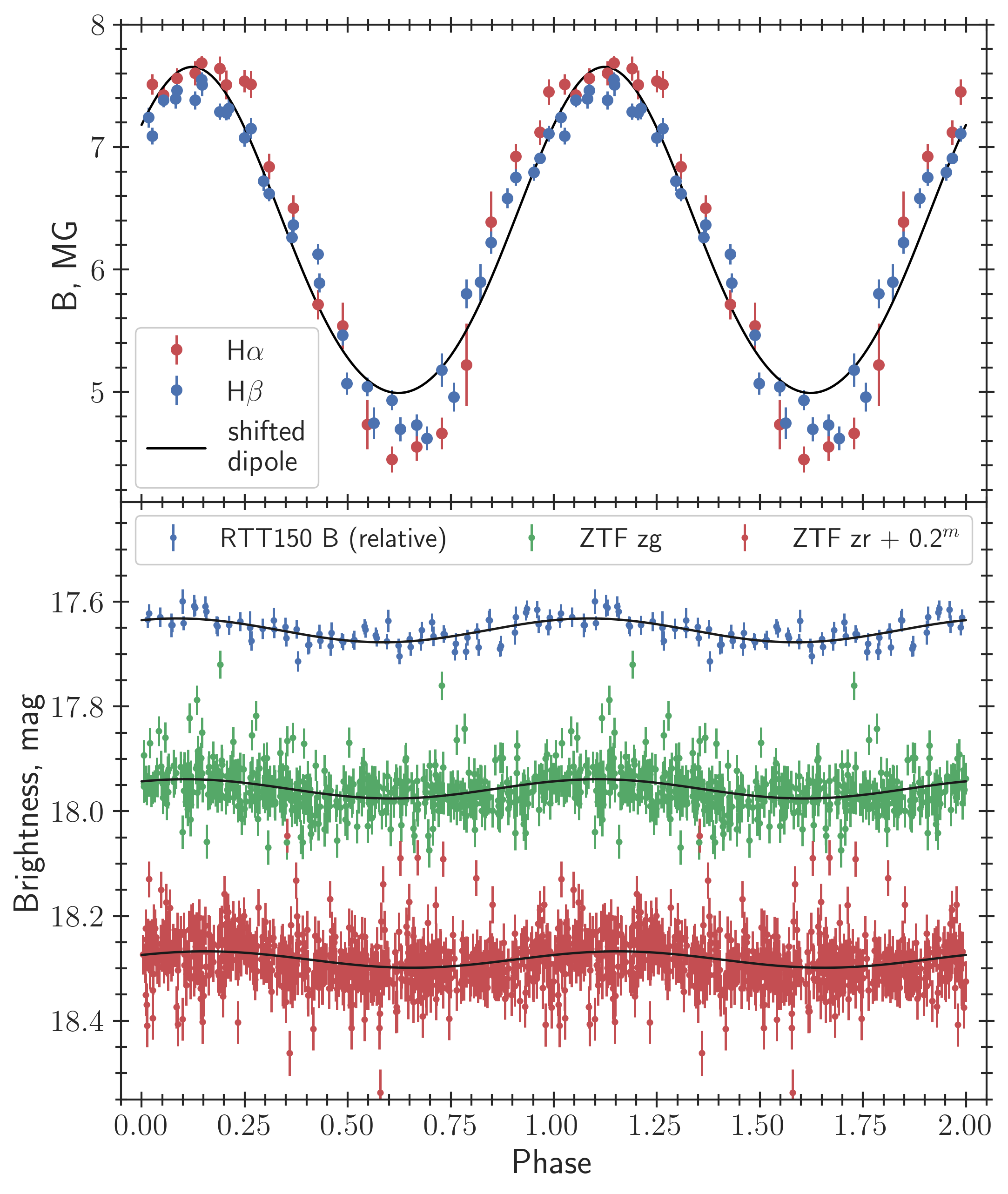}
    \vspace*{-5mm}
    \caption{Top panel: phased magnetic field curve of {\obj}, derived from the Zeeman splitting of the H$\alpha$ and H$\beta$ lines, with an offset dipole model overplotted. Bottom panel: optical light curves of {\obj} from ZTF and RTT-150 observations.}
    \label{fig:mag_curve}
    \end{minipage}
\end{figure}

Using the same approach, we determined the magnetic field from each spectrum in the two observational datasets. The resulting phased magnetic field curve over the orbital period is shown in the top panel of Fig.~\ref{fig:mag_curve}. The magnetic field exhibits a quasi-sinusoidal variation in the range $B \approx 4.5 - 7.5$~MG. Figure~\ref{fig:mag_curve} also shows that the strongest magnetic field occurs near the phase of maximum brightness in optical light curves, indicating accretion onto the region surrounding the white dwarf's magnetic pole.

\subsection{Magnetic field topology}
\label{Magnetic field topology}

The magnetic field configuration of the white dwarf was reconstructed using an offset dipole approximation. This model is defined by the following set of parameters: $i$ --- the inclination of the white dwarf's rotation axis to the line of sight, $\beta$ --- the tilt of the dipole axis relative to the rotation axis, $\psi$ --- the longitude of the magnetic pole measured from the direction toward the donor, $B_0$ --- the magnetic field strength at the magnetic pole, and $a$ --- the offset of the dipole center relative to the white dwarf's center (in units of $R_{\mathrm{wd}}$). The white dwarf's surface is divided into $n$ small segments, each assigned a local magnetic field calculated as
\begin{equation}
    B = \frac{B_0}{2} \bigg( \frac{1-a}{r} \bigg)^3 \big(3 \cos^2 \theta + 1 \big)^{1/2},
    \label{m_field}
\end{equation}
where $\theta$ is the polar angle of the surface element, and $r = \sqrt{1 + a^2 - 2 a \cos \theta}$ is the distance from the center of the element to the center of the dipole (in units of $R_{\mathrm{wd}}$).

The observed magnetic field $\bar{B}$ at a given rotational phase $\varphi$ is calculated by integrating the local magnetic field over the visible surface of the white dwarf:
\begin{equation}
    \bar{B}(\varphi) = \frac{\displaystyle\sum_{i=1}^{n} B_i S_i L_i(\varphi) \cos\gamma_i(\varphi)}{\displaystyle\sum_{i=1}^{n} S_i L_i (\varphi) \cos\gamma_i(\varphi)},
    \label{mean_mag}
\end{equation}
where the index $i$ denotes a surface patch, $\gamma_i$ is the angle between the patch normal and the line of sight at phase $\varphi$, $S_i$ is the surface area of the patch ($S_i = 0$ if $\cos \gamma_i < 0$), $B_i$ is the local magnetic field, and $L_i = 1 - c (1 - \cos \gamma_i)$ is a linear limb-darkening function. The limb-darkening coefficient was adopted as $c = 0.38$, appropriate for a white dwarf atmosphere with $T_{\mathrm{eff}} = 11000$~K, $\log g = 8.0$, and a wavelength of $\sim 470$~nm \citep{Gianninas2013}. The model magnetic curve was fitted to the observed data using a least-squares method in combination with the Nelder--Mead optimization algorithm.

\begin{figure}[h!]
    \centering
    \begin{minipage}[t]{\columnwidth}
    \centering
    \includegraphics[width=\linewidth]{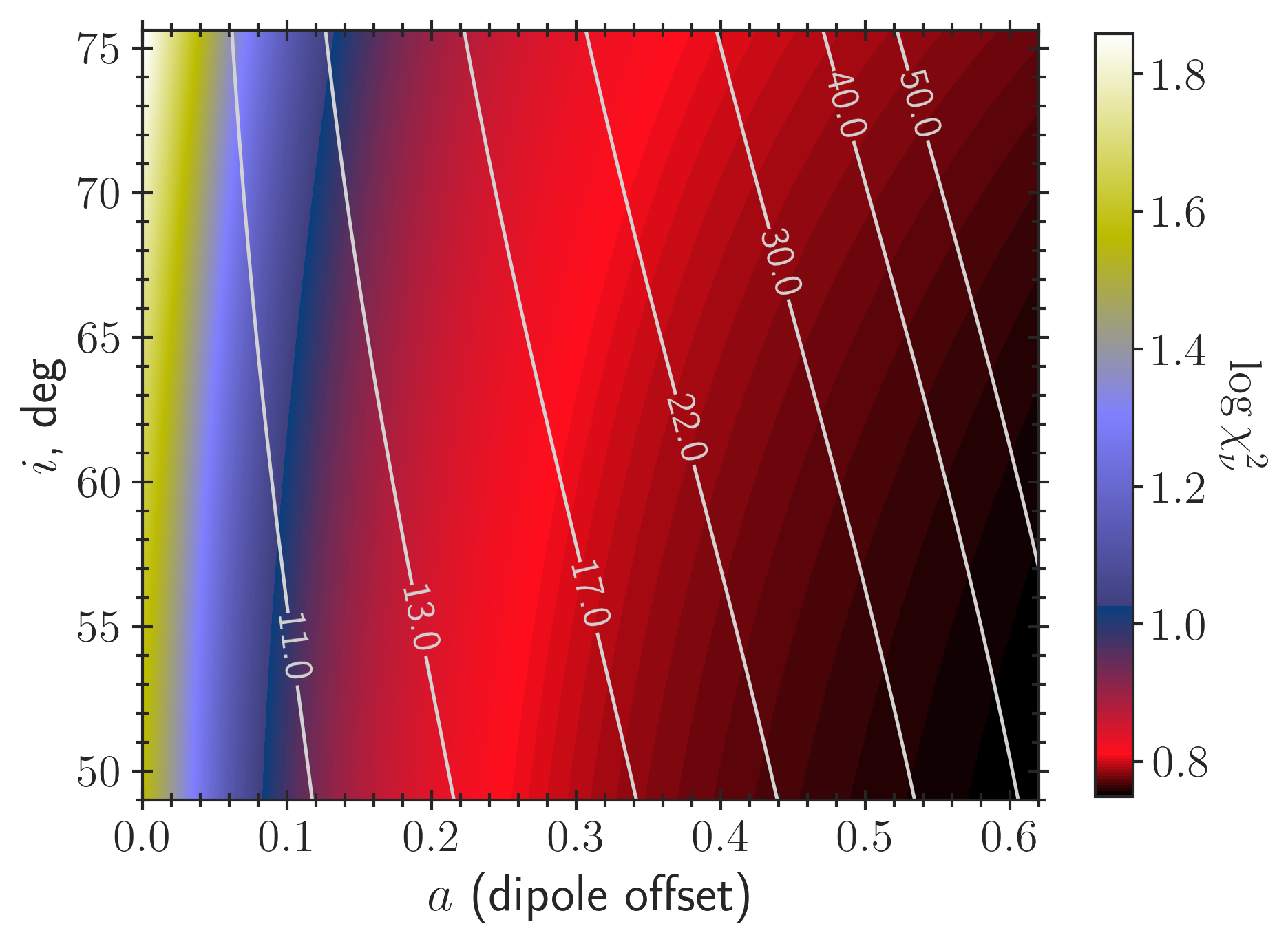}
    \vspace*{-5mm}
    \caption{Map of the reduced chi-square $\log \chi^2_{\nu}$ in the $i-a$ plane. The contours show the magnetic field strength $B_0$ at the magnetic pole.}
    \label{fig:chi_sq}
    \end{minipage}
\end{figure}

Figure~\ref{fig:chi_sq} shows the reduced chi-square distribution in the $i-a$ plane. The inclination angle $i$ was varied between $50$ and $80^{\circ}$. The lower limit of $i$ is derived from the assumption that the H$\alpha$ emission forms on the donor's surface, while the upper limit is constrained by the absence of eclipses of the white dwarf. For each pair of ($i$, $a$), the orientation of the magnetic dipole (angles $\beta$ and $\psi$) and the magnetic field strength $B_0$ were determined. The map also shows contours of the derived polar magnetic field $B_0$. It is apparent from the figure that reproducing the observed magnetic curve satisfactorily requires large offsets of the magnetic dipole, which correspond to polar magnetic field strengths of several tens of megagauss. However, such high values of $B_0$ are likely unrealistic given the relatively low magnetic field in the accretion spot ($B_{\rm spot} \approx 6$–$7$~MG, \citealt{Schmidt2005, Suslikov2025}) located near the magnetic pole.

\begin{figure}[h!]
    \centering
    \begin{minipage}[t]{\columnwidth}
    \centering
    \includegraphics[width=0.9\linewidth]{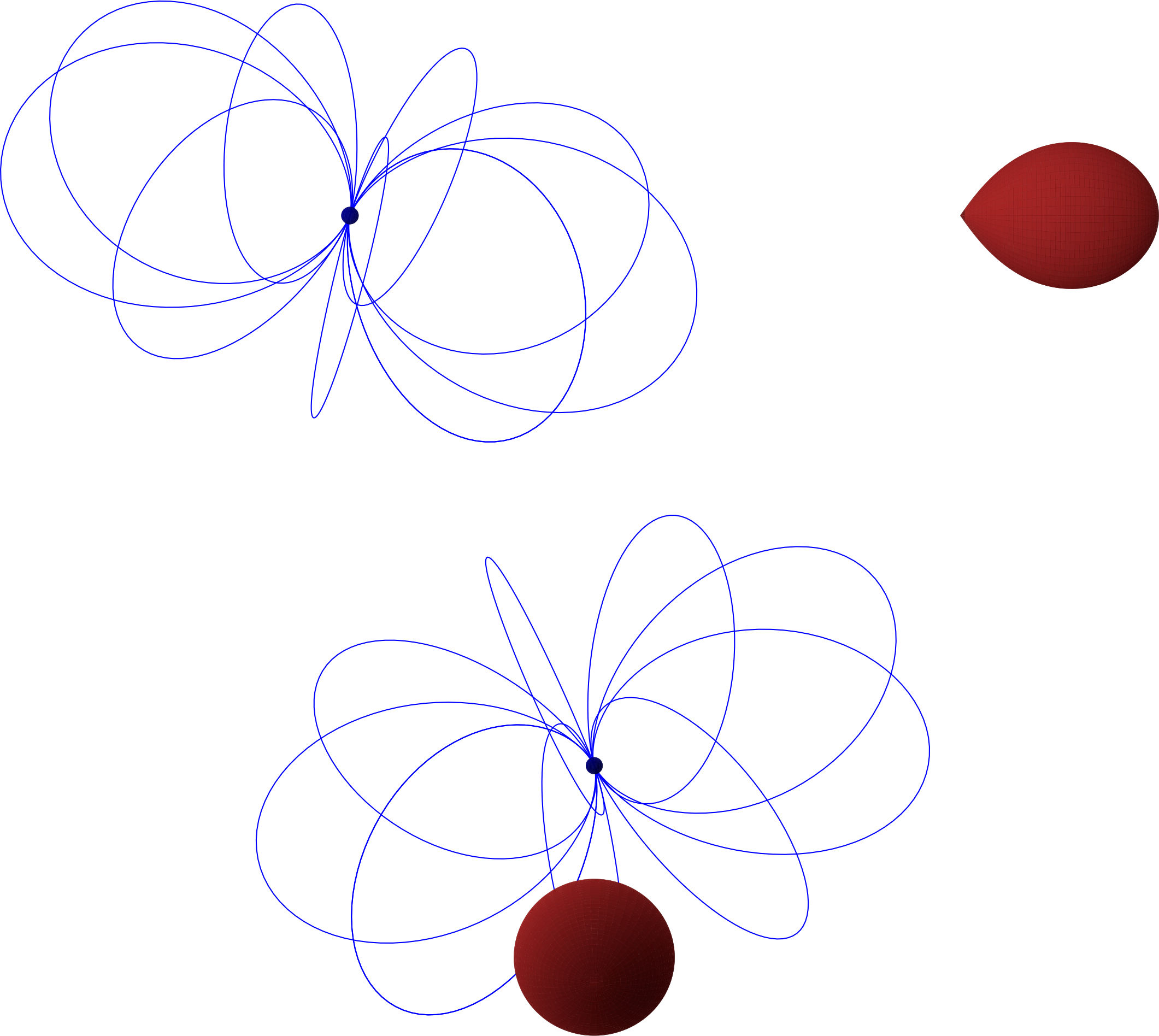}
    \vspace*{5mm}
    \caption{Magnetic field configuration of the white dwarf relative to the donor in the {\obj} binary model. The system is shown at orbital phases $\varphi=0$ (bottom) and $\varphi=0.25$ (top).}
    \label{fig:binary}
    \end{minipage}
\end{figure}

Figure~\ref{fig:mag_curve} shows the best-fit magnetic curve for an inclination of $i = 60^{\circ}$ and a polar magnetic field strength of $B_0 = 13$~MG~$\approx 2B_{\rm spot}$ ($a = 0.17$). The orientation of the magnetic dipole is given by the angles $\beta = 23.8^{\circ}$ and $\psi = 45.9^{\circ}$. It is apparent that the offset dipole model does not fully reproduce the minimum of the magnetic curve. The geometric model of the system, obtained for the given dipole orientation and the component parameters from \cite{Suslikov2025}, is shown in Fig.~\ref{fig:binary}.

Figure~\ref{fig:mag_curve} presents the best-fit magnetic curve for an inclination of $i = 60^{\circ}$ and a polar magnetic field strength of $B_0 = 13$~MG~$\approx 2B_{\rm spot}$ ($a = 0.17$). The orientation of the magnetic dipole is defined by the angles $\beta = 23.8^{\circ}$ and $\psi = 45.9^{\circ}$. It can be seen that the offset dipole model does not perfectly fit the minimum of the magnetic curve. The corresponding geometric model of the binary system, constructed using this dipole orientation and the component parameters from \cite{Suslikov2025}, is shown in Fig.~\ref{fig:binary}.

\section{CONCLUSION}
\label{Conslusion}

We have performed an analysis of phase-resolved spectroscopy of the magnetic period bouncer {\obj}. Examination of the dynamic spectra and Doppler tomography in the H$\alpha$ line indicates that most of the emission flux is likely produced in the accretion flow rather than on the surface of the donor star. For the first time, a magnetic curve of {\obj} has been derived, revealing rotational variability of the observed magnetic field in the range $4.5-7.5$~MG. Modeling of the magnetic curve suggests a complex magnetic field distribution on the white dwarf's surface, deviating from a simple dipolar configuration.

To confirm the hypothesis on the origin of the H$\alpha$ emission, phase-resolved spectroscopy with higher spectral resolution and a high signal-to-noise ratio is necessary. Understanding the nature of the H$\alpha$ line is essential for accurately determining the component masses in magnetic period bouncers, given the faintness of the secondary star and the challenges in modeling the spectra of magnetic white dwarfs. {\obj} represents a particularly valuable system for investigating the magnetic field topology of white dwarfs in close binary systems. Further study using spectropolarimetric observations would allow Zeeman-Doppler imaging of the white dwarf, providing deeper insights into its magnetic structure.

\section*{FUNDING}

Observations with the SAO RAS telescopes are supported by the Ministry of Science and Higher Education of the Russian Federation. The modernization of the observatory's equipment is carried out under the national project «Science and universities».

\section*{CONFLICT OF INTEREST}
The authors declare no conflicts of interest.

\bibliographystyle{aspb1}
\bibliography{main}

\end{document}